Astronomy
&
Astrophysics

LETTER TO THE EDITOR

# Resolved near-UV hydrogen emission lines at 40-Myr super-Jovian protoplanet Delorme 1 (AB)b

## Indications of magnetospheric accretion[⋆]

Simon C. Ringqvist[1], Gayathri Viswanath[1], Yuhiko Aoyama[2,3], Markus Janson[1],
Gabriel-Dominique Marleau[4,5,6,7], and Alexis Brandeker[1]

[1] Institutionen för Astronomi, Stockholms Universitetscentrum, 106 91 Stockholm, Sweden
e-mail: simon.ringqvist@astro.su.se
[2] Kavli Institute for Astronomy and Astrophysics, Peking University, Beijing 100871, PR China
[3] Institute for Advanced Study, Tsinghua University, Beijing 100084, PR China
[4] Fakultät für Physik, Universität Duisburg-Essen, Lotharstraße 1, 47057 Duisburg, Germany
[5] Institut für Astronomie und Astrophysik, Universität Tübingen, Auf der Morgenstelle 10, 72076 Tübingen, Germany
[6] Physikalisches Institut, Universität Bern, Gesellschaftsstr. 6, 3012 Bern, Switzerland
[7] Max-Planck-Institut für Astronomie, Königstuhl 17, 69117 Heidelberg, Germany

## ABSTRACT

*Context.* Accretion at planetary-mass companions (PMCs) suggests the presence of a protoplanetary disc in the system, likely accompanied by a circumplanetary disc. High-resolution spectroscopy of accreting PMCs is very difficult due to their proximity to bright host stars. For well-separated companions, however, such spectra are feasible and they are unique windows into accretion.
*Aims.* We have followed up on our observations of the 40-Myr, and still accreting, circumbinary PMC Delorme 1 (AB)b. We used high-resolution spectroscopy to characterise the accretion process further by accessing the wealth of emission lines in the near-UV.
*Methods.* We have used the UVES spectrograph on the ESO VLT/UT2 to obtain $R \approx 50\,000$ spectroscopy, at 3300–4520 Å, of Delorme 1 (AB)b. After separating the emission of the companion from that of the M5 low-mass binary, we performed a detailed emission-line analysis, which included planetary accretion shock modelling.
*Results.* We reaffirm ongoing accretion in Delorme 1 (AB)b and report the first detections in a (super-Jovian) protoplanet of resolved hydrogen line emission in the near-UV (Hγ, Hδ, Hε, H8, and H9). We tentatively detect H11, H12, He I, and Ca II H/K. The analysis strongly favours a planetary accretion shock with a line-luminosity-based accretion rate of $\dot{M} = 2 \times 10^{-8}\,M_J\,yr^{-1}$. The lines are asymmetric and are well described by sums of narrow and broad components with different velocity shifts. The overall line shapes are best explained by a pre-shock velocity of $v_0 = 170 \pm 30\,km\,s^{-1}$, implying a planetary mass of $M_P = 13 \pm 5\,M_J$, and number densities of $n_0 \gtrsim 10^{13}\,cm^{-3}$ or $n_0 \sim 10^{11}\,cm^{-3}$. The higher density implies a small line-emitting area of ∼1% relative to the planetary surface. This favours magnetospheric accretion, a case potentially strengthened by the presence of blueshifted emission in the line profiles.
*Conclusions.* High-resolution spectroscopy offers the opportunity to resolve line profiles, which are crucial for studying the accretion process in depth. The super-Jovian protoplanet Delorme 1 (AB)b is still accreting at ∼40 Myr. Thus, Delorme 1 belongs to the growing family of 'Peter Pan disc' systems with (a) protoplanetary and/or circumplanetary disc(s) far beyond the typically assumed disc lifetimes. Further observations of this benchmark companion and its presumed disc(s) will help answer key questions about the accretion geometry in PMCs.

## 1. Introduction

The process of mass accretion in young stars has been well studied and modelled for decades. Currently, magnetospheric accretion models best fit the observations (e.g., Hartmann et al. 2016). Matter in accretion columns is funnelled by magnetic fields onto the stellar surface and impacts at near free-fall velocities, releasing its kinetic energy at a shock. This leads to observable emission in a number of lines, most strongly in H I (e.g., Muzerolle et al. 1998; Herczeg & Hillenbrand 2008), which is often accompanied by He I and Ca II emission. The line luminosities can be used to estimate mass accretion rates (e.g., Rigliaco et al. 2012; Alcalá et al. 2017) and to distinguish accretion from chromospheric activity (e.g., Manara et al. 2017; Venuti et al. 2019). Line characteristics such as width (e.g., Jayawardhana et al. 2003; Natta et al. 2004) and profile asymmetries (e.g., Thanathibodee et al. 2022) can also aid in characterising the accretion process. Finally, the time variability of stellar accretion has also been studied using high-resolution spectroscopy, for example in EX Lupi during both outbursts and quiescence (Sicilia-Aguilar et al. 2012, 2015). EX Lupi showed strong variability in nearly all observed emission lines, including numerous metallic emission lines in the near-UV such as Fe I, Fe II, and Ti II.

Contrary to the stellar case, accretion in brown dwarfs or planetary-mass companions (PMCs) has only more recently, thanks to advancements in instrumentation and post-processing techniques, become accessible for study (e.g., Haffert et al. 2019; Eriksson et al. 2020; Stolker et al. 2021; Currie et al. 2022; Zhou et al. 2022). The most widely studied of these are the two protoplanets in the 5 Myr PDS 70 system (Keppler et al. 2018; Müller et al. 2018). Atacama Large Millimeter Array (ALMA) observations also confirmed the presence of at least one circumplanetary disc (CPD) in the system (Christiaens et al. 2019; Isella et al. 2019; Benisty et al. 2021). Since the detection of Hα emission from both PDS 70 b and c or their respective CPDs by Haffert et al. (2019), accretion in these protoplanets has been characterised and modelled in a number of works (e.g., Thanathibodee et al. 2019; Aoyama & Ikoma 2019; Hashimoto et al. 2020; Wang et al. 2020, 2021; Uyama et al. 2021; Zhou et al. 2021). So far, emission lines other than Hα have not been detected from either source, and observational constraints have only allowed for medium-resolution spectroscopy ($R \lesssim 5000$). At this resolution, the line widths cannot be reliably measured (e.g., Thanathibodee et al. 2019; Eriksson et al. 2020), and line asymmetries that can further constrain the source of the emission (e.g., shocks onto the surface of the planet or the CPD), and therefore the accretion geometry, are smoothed out (e.g., Thanathibodee et al. 2022; Marleau et al. 2022). In general, the observed signatures of accretion in the planetary paradigm can be explained by a similar magnetospheric model as in the stellar accretion paradigm (e.g., Thanathibodee et al. 2019), but also by other models (e.g., Aoyama et al. 2018, 2020; Marleau et al. 2019, 2022; Szulágyi & Ercolano 2020; Takasao et al. 2021). Determining which one best describes the accretion process in PMCs requires the use of high-resolution spectroscopy.

In Eriksson et al. (2020), we observed the 12–14 $M_J$ PMC Delorme 1 (AB)b (or 2MASS J01033563−5515561 (AB)b; Delorme et al. 2013) with VLT/Multi-Unit Spectroscopic Explorer (MUSE). We discovered very strong Hα emission, as well as Hβ and He I emission, indicating accretion. The discovery was unexpected due to the ∼40 Myr age of the system, based on its Tuc-Hor membership. This makes it an unusual and uniquely 'old' accreting PMC, but not unique when compared with stellar or brown dwarf accretors of similar ages, referred to as 'Peter Pan disc' systems (e.g., Boucher et al. 2016; Silverberg et al. 2016, 2020; Murphy et al. 2018; Lee et al. 2020). Betti et al. (2022) recently detected near-infrared H I emission (Paβ, Paγ, and Brγ) confirming the accretion status of Delorme 1 (AB)b. Due to being relatively nearby (47.2 ± 3.1 pc; Riedel et al. 2014) and at a ∼1.8″ separation from its faint M5 binary host, Delorme 1 (AB)b is uniquely suited for follow-up with high-resolution spectroscopy.

In this Letter we present the results (Sect. 3) of our near-UV (3300–4520 Å), $R_\lambda = \lambda/\Delta\lambda \approx 50\,000$ VLT/Ultraviolet and Visual Echelle Spectrograph (UVES) observations (ID: 0108.C-0655(A)) of Delorme 1 (AB)b, which offer a unique window into an accreting protoplanet. We report the detection of a large number of emission lines, primarily from H I (H5–H9 and H11–H12) where H5–H8 also show line asymmetries. The implications of these asymmetries are discussed further in Sect. 4.

## 2. UVES observations and data reduction

The Delorme 1 system was observed using UVES (Dekker et al. 2000), mounted on the ESO VLT/UT2 in Cerro Paranal, Chile. Depending on the instrumental setup, UVES provides a wave-length coverage of $\lambda \approx 3000$–11 000 Å, split into two arms (blue and red), with $R_\lambda \approx 40\,000$–11 0000. The UVES observations were obtained using a 0.8″ wide slit covering 3300–4520 Å in the blue arm at $R_\lambda \approx 50\,000$. Table A.1 provides the observing log.

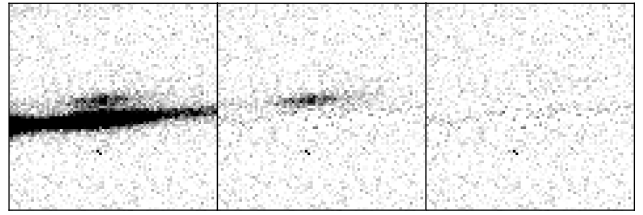

**Fig. 1.** Cut-out of frame 1 showing H7 in order 24. *Left*: before subtraction of the stellar lines. *Middle*: after subtraction. *Right*: residuals after subtraction, showing the efficient extraction of the companion line.

Data reduction and spectral extraction were done using a custom pipeline that specialises in separating the companion spectrum from that of the host star, through an iterative point spread function (PSF) fitting process (see Appendix A.1). Figure 1 shows an example of the extraction of H7 in order 24, with other lines in Figs. A.1–A.3.

The extracted spectrum was flux-calibrated and blaze-curvature-corrected order by order (see Appendix A.2). The spectrum is consistent between both frames, suggesting non-variability on very short timescales (∼0.5 h). These were therefore combined in an averaged spectrum, which is used for the analysis presented in Sect. 3. The flux uncertainty in a given spectral bin of the averaged spectrum was set to the standard deviation of a three-bin-wide moving box in each frame (six bins in total). For integrated absolute line fluxes, uncertainties from the flux calibration and blaze curvature correction dominate.

## 3. Analysis and results

We have resolved, for the first time in a protoplanet, H I line emission in the near-UV (H5–H9). For emission line characteristics, readers can refer to Tables 1 and B.1, as well as Figs. 2 and B.1–B.6. We also report a number of tentative detections, notably H11, H12, He I (4471.66, 4026.34, and 3888.64 Å), Ca II H and K (3968.47 and 3933.66 Å), as well as a number of lines that could correspond to neutral metals (Fe I, Cr I, and Ti I). Figure B.7 shows an example of the He I $\lambda$ 4471.66 detection, Table B.2 lists all tentative detections, and their characterisation is described in Appendix B.2.

### 3.1. Characterising the near-UV H I line emission

Following the steps outlined in Sect. 2, a barycentric correction was applied and calculated to −13.234 km s⁻¹ using `astropy.skycoord` at Cerro Paranal on 2021 October 25. Unambiguous detections were made of H I emission lines from H5 (Hγ, 4340.464 Å) down to H9 (3835.397 Å). We tentatively detected H11 and H12, while H10 was not due to a low signal-to-noise ratio (S/N). Table 1 provides integrated line fluxes ($F_{line}$) and luminosities ($L_{line}$). We note that H6 might be overluminous in this context, being brighter than H5. Similarly, there is a discrepancy for the absolute fluxes of the two different orders of H7, where it seems likely that the true flux is closer to an average of the two. On the other hand, as shown in Fig. B.1, the line profiles of both orders agree well. Therefore, the discrepancies in flux most likely come from differences in the order by order flux calibration.

**Table 1.** Delorme 1 (AB)b integrated fluxes, luminosities, and derived mass accretion rates for detected H I emission lines.

| Line | $F_{line}$ | | Stellar scaling $^{(a)}$ | | Planetary scaling $^{(b)}$ | |
| | | $\log(L_{line})$ | $\log(L_{acc})$ | $\log(\dot{M})$ | $\log(L_{acc})$ | $\log(\dot{M})$ |
| | $(10^{-16}$ erg cm$^{-2}$ s$^{-1})$ | $(L_\odot)$ | $(L_\odot)$ | $(M_J$ yr$^{-1})$ | $(L_\odot)$ | $(M_J$ yr$^{-1})$ |
|---|---|---|---|---|---|---|
| H5$_{33}$ (Hγ) | $11.3^{+5.7}_{-3.8}$ | $-7.1 \pm 0.3$ | $-5.2 \pm 0.5$ | $-8.4 \pm 0.5$ | $-4.4 \pm 0.3$ | $-7.7 \pm 0.3$ |
| H6 (Hδ) | $18.0^{+3.6}_{-3.0}$ | $-6.9 \pm 0.1$ | $-4.7 \pm 0.4$ | $-8.0 \pm 0.4$ | $-4.0 \pm 0.2$ | $-7.3 \pm 0.2$ |
| H7$_{24}$ (Hε) | $7.1^{+1.4}_{-1.2}$ | $-7.3 \pm 0.1$ | $-5.1 \pm 0.4$ | $-8.3 \pm 0.4$ | $-4.2 \pm 0.2$ | $-7.4 \pm 0.2$ |
| H7$_{23}$ (Hε) | $1.6^{+1.0}_{-0.6}$ | $-7.9 \pm 0.3$ | $-5.7 \pm 0.5$ | $-9.0 \pm 0.5$ | $-4.7 \pm 0.3$ | $-7.9 \pm 0.3$ |
| H8 | $2.2^{+0.9}_{-0.6}$ | $-7.8 \pm 0.2$ | $-5.6 \pm 0.4$ | $-8.8 \pm 0.4$ | $-4.5 \pm 0.3$ | $-7.7 \pm 0.3$ |
| H9 | $1.1^{+0.5}_{-0.3}$ | $-8.1 \pm 0.2$ | $-5.7 \pm 0.4$ | $-8.9 \pm 0.4$ | $-4.6 \pm 0.3$ | $-7.8 \pm 0.3$ |
| H11 | $0.5^{+0.4}_{-0.3}$ | $-8.4 \pm 0.4$ | $-5.9 \pm 0.5$ | $-9.2 \pm 0.5$ | $-4.6 \pm 0.4$ | $-7.8 \pm 0.4$ |
| H12 | $0.4^{+0.3}_{-0.2}$ | $-8.5 \pm 0.4$ | $-5.8 \pm 0.5$ | $-9.1 \pm 0.5$ | $-4.6 \pm 0.4$ | $-7.8 \pm 0.4$ |

**Notes.** Line subscripts denote the spectral order. Integrated fluxes are extinction-corrected. Lines H11–H12 are tentative (Appendix B.2). $^{(a)}$ $L_{acc}$ derived from $L_{line}$ using the stellar scaling relations in Alcalá et al. (2017). $^{(b)}$ $L_{acc}$ derived from $L_{line}$ using the planetary scaling relations in Aoyama et al. (2021) for H5–H8, and in Marleau & Aoyama (2022) for H9, $(a, b) = (0.83, 2.16)$; H11, $(0.82, 2.37)$; and H12, $(0.82, 2.47)$.

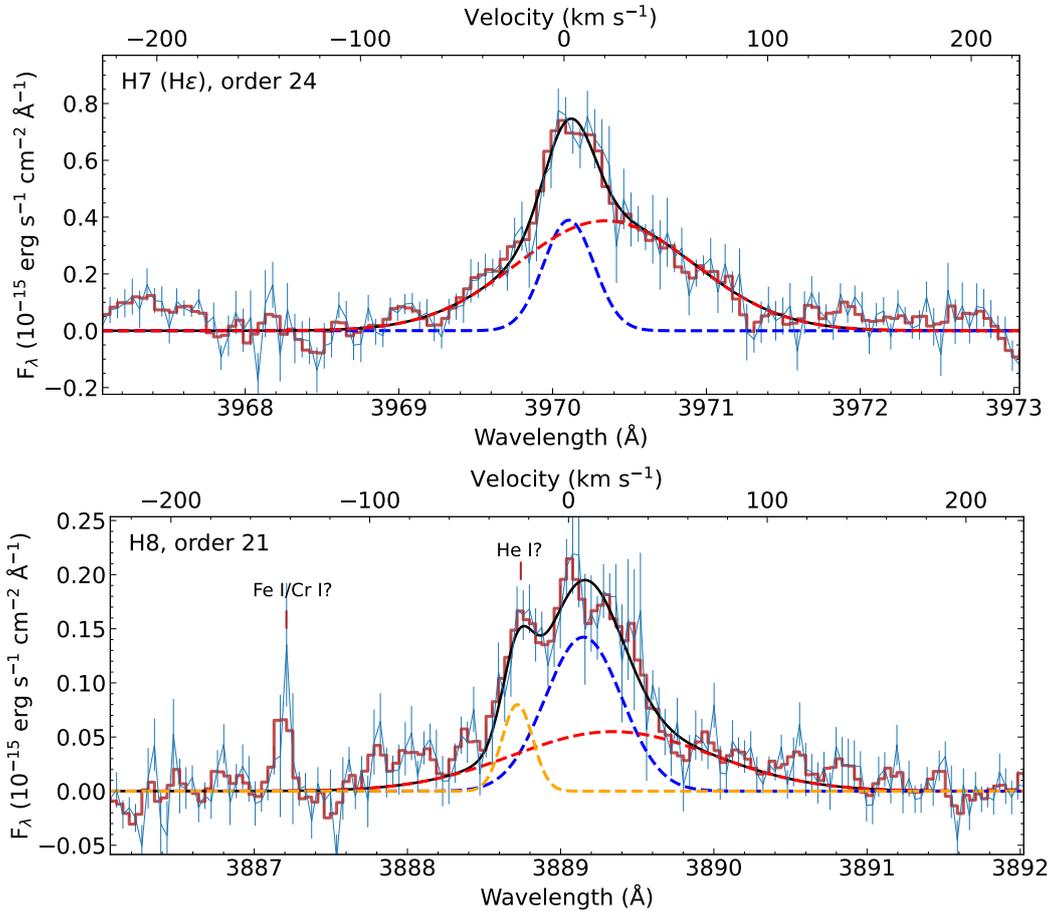

**Fig. 2.** H7 and H8 line profiles (light blue; light red: three-bin moving average). Clear asymmetries in both lines preclude single-Gaussian fits. The fit (black line) is the sum of narrow and broad components (NC, dashed blue; BC, red), and might also show Fe I 3887.05 Å or Cr I 3887.08 Å emission. The velocity zero-point is set to the rest-frame wavelength of the H I emission line.

In the stellar accretion paradigm, scaling relations have been derived that correlate $L_{line}$ with an accretion luminosity ($L_{acc}$) (Rigliaco et al. 2012; Alcalá et al. 2017), according to

$$\log(L_{acc}/L_\odot) = a \times \log(L_{line}/L_\odot) + b \quad (1)$$

where $a$ and $b$ correspond to the fit coefficients of the scaling relation for a given line. Mass accretion rates $\dot{M}$ are then

derived

$$\dot{M} = \left(1 - \frac{R_\star}{R_{in}}\right)^{-1} \frac{L_{acc} R_\star}{G M_\star}, \quad (2)$$

where $R_\star$ is the radius of the accreting object, $R_{in}$ is the inner disc radius, conventionally set to $5\,R_\star$ (e.g., Gullbring et al. 1998; Herczeg & Hillenbrand 2008; Alcalá et al. 2017), and $M_\star$ is the mass of the object. Based on the isochronal analysis in Eriksson et al. (2020), we adopted $M_P = 0.012\,M_\odot$ and $R_P = 0.163 \pm 0.009\,R_\odot$, and used the coefficients derived by Alcalá et al. (2017) to calculate $L_{acc}$ and $\dot{M}$ for H5–H9 and H11–H12. The uncertainty in $L_{acc}$ is dominated by the scatter in the scaling relation, rather than the uncertainty in $L_{line}$.

In Eriksson et al. (2020), we also noted a large discrepancy between mass accretion rates estimated by the stellar scaling relation and those estimated by planetary accretion models. The work of Aoyama et al. (2021) proposed a new planetary scaling relation, formulated in the same way as Eq. (1). It is based on detailed non-equilibrium shock calculations with input parameters (velocity and number density) appropriate for forming planets (see details in Aoyama et al. 2018, 2020). Using the corresponding $(a, b)$ coefficients for H5–H8, with H9–H12 from Marleau & Aoyama (2022), we present planetary scaling estimates for $L_{acc}$ and $\dot{M}$ in Table 1, along with the stellar scaling estimates. On average, we find $\log(\dot{M}/M_J\,yr^{-1}) = -8.8 \pm 0.4$ from the stellar scaling of Alcalá et al. (2017), and $\log(\dot{M}/M_J\,yr^{-1}) = -7.7 \pm 0.3$ from the planetary scaling of Aoyama et al. (2021). Revisiting the 2018 September 18 MUSE epoch in Eriksson et al. (2020) and using this planetary scaling for H$\alpha$ and H$\beta$, we find an average $\log(\dot{M}/M_J\,yr^{-1}) = -8.5 \pm 0.1$. Closest in time to the UVES epoch is the 2021 November 20 epoch near-infrared (NIR) spectrum by Betti et al. (2022), for which they find an average $\log(\dot{M}/M_J\,yr^{-1}) = -7.4 \pm 0.2$.

### 3.2. Characterising line profile asymmetries

Several of the H I lines display clear asymmetries, which are primarily redshifted and especially prominent in H7 and H8 (Fig. 2). This suggests that the lines are composed of both narrow (NC) and broad (BC) components, similar to what Sicilia-Aguilar et al. (2012, 2015) found for the accreting star EX Lupi and as predicted by the planetary accretion shock framework of Aoyama et al. (2021). In stellar magnetospheric accretion, the NC originates in the post-shock region, while the BC likely originates in the pre-shock flow (Hartmann et al. 2016). In the planetary accretion shock framework, the gas in the immediate post-shock region is hotter ($T \gtrsim 10^5$ K) and is still receding quickly away from the observer, down into the planet. This results in a redshift of the BC relative to the NC, and a broader profile width due to greater Doppler broadening. The NC in turn originates deep below the shock at lower temperatures ($T \approx 10^4$ K) and is expected to have a full width at half maximum (FWHM) of a few tens of km s$^{-1}$ (Aoyama et al. 2020). To disentangle these components, all lines were analysed through an iterative fitting procedure (see Appendix B.1), which included both single and multiple Gaussians. Figure 2 shows the best fits for H7$_{24}$ and H8 (with the rest in Figs. B.1–B.6), and Table B.1 provides the fit parameters. Section 4 discusses the implications of these findings.

Line profiles were separately analysed within the framework of Aoyama et al. (2020, 2021). Here, each line profile is compared to a model grid with pre-shock number densities of $n_0 = 10^9$–$10^{14}$ cm$^{-3}$ and pre-shock velocities of $v_0 = 50$–$200$ km s$^{-1}$. Figure B.8 shows the normalised root mean square error (RMSE) for the accretion shock model. These were created by comparing the observed line profiles with the model as a function of $(n_0, v_0)$. The RMSE measures the quality of the fit by normalising the deviation by the noise dispersion ($\sigma_{F_{cont}}$), namely,

$$\text{RMSE} = \sqrt{\frac{1}{N} \frac{\sum (F_{line} - F_{model})^2}{\sigma_{F_{cont}}^2}}, \tag{3}$$

where $F_{model}$ is the modelled line flux and $N$ is the number of data points. With no detected continuum, the average $F_{cont}$ is close to zero. The red (blue) region denotes where the model–observation discrepancy is larger (smaller) than the noise level.

How much the grid parameters ($n_0, v_0$) can be constrained depends on (i) the line-S/N and (ii) the sensitivity of the line profile on the grid parameters. The line profile becomes wider with increasing $v_0$, but it hardly depends on $n_0$, except when the line becomes optically thick and saturated (Aoyama & Ikoma 2019). Since the hydrogen excitation originates from collisional excitation in the post-shock, the more highly excited hydrogen is less abundant and optically thinner. Consequently, the less excited hydrogen (e.g., source of H5 or H6) is more sensitive to $n_0$.

This analysis and Fig. B.8 indicate a best-fit pre-shock velocity of $v_0 = 170 \pm 30$ km s$^{-1}$. For $n_0$, two different ranges are indicated, $n_0 \sim 10^{11}$ cm$^{-3}$ and $n_0 \gtrsim 10^{13}$ cm$^{-3}$. At the lower $n_0$, the BC of the model matches only the NC of the data. At the higher $n_0$, the model reproduces both the NC and BC of the data. Assuming the pre-shock velocity to be a free-fall velocity yields an estimate of the accretor's mass $M_P$ from $v_0^2 R_P = 2GM_P$. This results in $M_P = 13 \pm 5\,M_J$ (uncertainty dominated by that on $v_0$), which agrees well with the isochronal masses from Delorme et al. (2013) and Eriksson et al. (2020).

From $\dot{M}$, $v_0$, and $n_0$, we can estimate the filling factor $f_{fill}$, the line-emitting fraction of the planetary surface area. By definition (Aoyama & Ikoma 2019), with $\mu$ a mean molecular weight,

$$\dot{M} = 4\pi R_P^2 n_0 v_0 \mu f_{fill}. \tag{4}$$

Equation (4) implies $f_{fill} \lesssim 0.3\%$ for $n_0 \gtrsim 10^{13}$ cm$^{-3}$ and $f_{fill} \sim 33\%$ for $n_0 \sim 10^{11}$ cm$^{-3}$.

An analysis was also done for the line flux ratios within the same framework. The results were inconclusive for H5 and H6, due to the large uncertainties in $F_{line}$. The line ratios between H7–H9 agree with the model prediction with any model parameters. This is because the H7–H9 ratios hardly depend on density or temperature (set by $v_0$), due to the short collisional-transition timescale between highly excited electron levels. Therefore, unlike the lines analysed in Betti et al. (2022), we cannot put constraints on $n_0$ or $v_0$ from line ratios of the higher-order lines.

## 4. Discussion

We have presented $R_\lambda \approx 50\,000$ spectroscopic observations of Delorme 1 (AB)b from the blue arm of VLT/UVES (see Viswanath et al., in prep. for the red arm). They provide the first detection of resolved H I line emission in the near-UV at a protoplanet. We now discuss a few points beyond the scope of the analysis of the H I lines presented in Sect. 3.

Using the Aoyama et al. (2021) planetary scaling relations and averaging the accretion rates across all H I lines yields an estimate of $\dot{M} = 2 \times 10^{-8}\,M_J\,yr^{-1}$ for Delorme 1 (AB)b at this epoch (Table 1). This agrees well with the NIR-based estimate of $\dot{M} = (3–4) \times 10^{-8}\,M_J\,yr^{-1}$ of Betti et al. (2022), who used the same scaling relations and non-contemporaneous observations obtained within 1–5 months of the UVES epoch. Comparing these two data sets with the original MUSE detection of accretion in Eriksson et al. (2020), it is possible that accretion in Delorme 1 (AB)b is variable by roughly an order of magnitude on a timescale of years.

Resolved line profiles allow for an analysis beyond line fluxes and probe possible accretion geometries. As noted in Sect. 3.2, the accretion flow likely has a pre-shock velocity of $v_0 = 170 \pm 30$ km s$^{-1}$ and slightly favours a number density of $n_0 \gtrsim$ $10^{13}$ cm$^{-3}$, as the model at higher $n_0$ reproduces both the observed NC and BC. Since $v_0$ and $\dot{M}$ are constrained within $\approx 30\%$, it is meaningful to discuss the filling fraction. The estimate of $f_{\text{fill}} \sim 0.3\%$ points more towards a magnetospheric accretion geometry. A spherical accretion geometry requires $f_{\text{fill}} \sim 1$ and should be ruled out at this $n_0$, but not at $n_0 \sim 10^{11}$ cm$^{-3}$. The higher $n_0$ is consistent with the estimate from Betti et al. (2022).

The multi-component Gaussian fits in Table B.1 favour NC$_{\text{FWHM}} \approx 40$ km s$^{-1}$ and BC$_{\text{FWHM}} \approx 90$–160 km s$^{-1}$ (for redshifted BCs). The widths and radial velocity (RV) shifts agree well with an origin at a planetary accretion shock (Sect. 3.2). Higher-order lines (H7 and above) are also predicted to show a stronger redshifted BC than less excited lines such as H5–H6 (Aoyama et al. 2020), which can be seen by comparing H7$_{24}$ (Fig. 2) with H5 (Fig. B.2), for example. For the two-component fits, there is a potential discrepancy in the RV shifts in Table B.1. For H5–H6, both the NC and BC are located at similar shifts ($\approx 7$–8 km s$^{-1}$); whereas, for H7–H8, the NC is located at $\approx 3$ km s$^{-1}$, while the BC is shifted by $\approx 20$ km s$^{-1}$. One interesting explanation could be that there is actually both a (slightly narrower) redshifted BC from the post-shock and a blueshifted BC from the pre-shock gas. The three-component fits of H7$_{24}$ and H6 in Table B.1 were obtained to explore this possibility (see Fig. B.4 for H6). These fits, while subject to stronger degeneracies, provide a better overlap and indicate similar velocities for the blueshifted components.

Blueshifted emission of this type is common in stellar magnetospheric accretion models where it originates in accretion funnels. The heating mechanism of these funnels remains unclear, but it is conceivable for magnetospheric accretion at planetary masses to lead to such emission (Thanathibodee et al. 2019). This would be a natural explanation since these blueshifted features are not expected from (planetary) accretion shocks and thus could suggest magnetospheric accretion in Delorme 1 (AB)b.

Higher-S/N spectra, or averages of contemporaneous observations, would help to reveal the true nature of these blueshifted features and to confirm the presence of metal lines which have only tentatively been detected (Table B.2). This would be beneficial as metal transitions of similar energy can be used to probe the local conditions of the gas, independently of hydrogen transitions (e.g., Sicilia-Aguilar et al. 2015). Similar as for Ex Lupi, such observations could examine the geometry of accretion shock hotspots on the surface, if done within days of each other, by looking for RV modulations.

In summary, Eriksson et al. (2020), Betti et al. (2022), and now this work show clearly that the super-Jupiter Delorme 1 (AB)b is still accreting despite its 'advanced' age of 40 Myr. This puts it at tension with the canonical estimate of much shorter protoplanetary disc lifetimes. As we have noted in Eriksson et al. (2020), this might suggest that the Delorme 1 system hosts a long-lived 'Peter Pan disc'. Disc simulations attempting to obtain these unusually long-lived discs (e.g., Coleman & Haworth 2020, 2022; Wilhelm & Zwart 2021) list a high disc mass ($M_{\text{disc}}/M_{\star} > 0.25$) as being key. Haworth et al. (2020) also find that low-mass stars can more easily support high disc masses, $M_{\text{disc}}/M_{\star} > 0.30$, than solar-type stars. Such a high disc mass also favours planet formation through gravitational instability (e.g., Mercer & Stamatellos 2020), which is a plausible scenario for Delorme 1 (AB)b given the architecture of the system. This reinforces the need to detect and characterise the gaseous disc(s) fuelling this accretion.

*Acknowledgements.* We thank the anonymous referee for the insightful comments that improved this Letter. MJ gratefully acknowledges support from the Knut and Alice Wallenberg Foundation, and G-DM from the DFG priority program SPP 1992 "Exploring the Diversity of Extrasolar Planets" (MA 9185/1) and from the Swiss National Science Foundation under grant 200021_204847 "PlanetsInTime". Parts of this work have been carried out within the framework of the NCCR PlanetS supported by the SNSF. This publication makes use of VOSA, developed under the Spanish Virtual Observatory project funded by MCIN/AEI/10.13039/501100011033/ through grant PID2020-112949GB-I00.

## References

Alcalá, J. M., Manara, C. F., Natta, A., et al. 2017, A&A, 600, A20
Aoyama, Y., & Ikoma, M. 2019, ApJ, 885, L29
Aoyama, Y., Ikoma, M., & Tanigawa, T. 2018, ApJ, 866, 84
Aoyama, Y., Marleau, G.-D., Mordasini, C., & Ikoma, M. 2020, MNRAS, 504, 888
Aoyama, Y., Marleau, G.-D., Ikoma, M., & Mordasini, C. 2021, ApJ, 917, L30
Astropy Collaboration (Price-Whelan, A., et al.) 2018, AJ, 156, 123
Bayo, A., Rodrigo, C., Barrado y Navascués, D., et al. 2008, A&A, 492, 277
Benisty, M., Bae, J., Facchini, S., et al. 2021, ApJ, 916, L2
Betti, S. K., Follette, K. B., Ward-Duong, K., et al. 2022, ApJ, 935, L18
Boucher, A., Lafrenière, D., Gagné, J., et al. 2016, ApJ, 832, 50
Christiaens, V., Cantalloube, F., Casassus, S., et al. 2019, ApJ, 877, L33
Coleman, G. A. L., & Haworth, T. J. 2020, MNRAS, 496, L111
Coleman, G. A. L., & Haworth, T. J. 2022, MNRAS, 514, 2315
Currie, T., Lawson, K., Schneider, G., et al. 2022, Nat. Astron., 6, 751
Dekker, R., D'Odorico, S., Kaufer, A., Delabre, B., & Kotzlowski, H. 2000, in Optical and IR Telescope Instrumentation and Detectors, SPIE, 4008, 534
Delorme, P., Gagné, J., Girard, J. H., et al. 2013, A&A, 553, L5
Earl, N., Tollerud, E., Jones, C., et al. 2022, astropy/specutils: V1.7.0
Eriksson, S. C., Asensio Torres, R., Janson, M., et al. 2020, A&A, 638, L6
Gullbring, E., Hartmann, L., Briceño, C., & Calvet, N. 1998, ApJ, 492, 323
Haffert, S. Y., Bohn, A. J., de Boer, J., et al. 2019, Nat. Astron., 3, 749
Hartmann, L., Herczeg, G., & Calvet, N. 2016, ARA&A, 54, 135
Hashimoto, J., Aoyama, Y., Konishi, M., et al. 2020, ApJ, 159, 222
Haworth, T. J., Cadman, J., Meru, F., et al. 2020, MNRAS, 494, 4130
Herczeg, G. J., & Hillenbrand, L. A. 2008, ApJ, 681, 594
Isella, A., Benisty, M., Teague, R., et al. 2019, ApJ, 879, L25
Jayawardhana, R., Mohanty, S., & Basri, G. 2003, ApJ, 592, 282
Keppler, M., Benisty, M., Müller, A., et al. 2018, A&A, 617, A44
Lee, J., Song, I., & Murphy, S. 2020, MNRAS, 494, 62
Leike, R. H., & Enßlin, T. A. 2019, A&A, 631, A32
Manara, C. F., Frasca, A., Alcalá, J. M., et al. 2017, A&A, 605, A86
Marleau, G.-D., & Cumming, A. 2014, MNRAS, 437, 1378
Marleau, G.-D., Aoyama, Y., et al. 2021, Res. Notes. Am. Astron. Soc., 6, 12
Marleau, G.-D., Mordasini, C., & Kuiper, R. 2019, ApJ, 881, 144
Marleau, G.-D., Aoyama, Y., Kuiper, R., et al. 2022, A&A, 657, A38
Mercer, A., & Stamatellos, D. 2020, A&A, 633, A116
Müller, A., Keppler, M., Henning, T., et al. 2018, A&A, 617, L2
Murphy, S. J., Mamajek, E. E., & Bell, C. P. M. 2018, MNRAS, 476, 3290
Muzerolle, J., Calvet, N., & Hartmann, L. 1998, ApJ, 492, 743
Natta, A., Testi, L., Muzerolle, J., et al. 2004, A&A, 424, 603
Riedel, A. R., Finch, C. T., Henry, T. J., et al. 2014, AJ, 147, 85
Rigliaco, E., Natta, A., Testi, L., et al. 2012, A&A, 548, A56
Robitaille, T. P., Tollerud, E. J., Greenfield, P., et al. 2013, A&A, 558, A33
Sicilia-Aguilar, A., Kóspál, Á., Setiawan, J., et al. 2012, A&A, 544, A93
Sicilia-Aguilar, A., Fang, M., Roccatagliata, V., et al. 2015, A&A, 580, A82
Silverberg, S. M., Kuchner, M. J., Wisniewski, J. P., et al. 2016, ApJ, 830, L28
Silverberg, S. M., Wisniewski, J. P., Kuchner, M. J., et al. 2020, ApJ, 890, 106
Squicciarini, V., & Bonavita, M. 2022, A&A, 666, A15
Stolker, T., Haffert, S. Y., Kesseli, A. Y., et al. 2021, AJ, 162, 286
Szulágyi, J., & Ercolano B. 2020, ApJ, 902, 126
Takasao, S., Aoyama, Y., & Ikoma, M. 2021, ApJ, 921, 10
Thanathibodee, T., Calvet, N., Bae, J., Muzerolle, J., & Hernández, R. F. 2019, ApJ, 885, 94
Thanathibodee, T., Calvet, N., Hernández, J., Maucó, K., & Briceño, C. 2022, AJ, 163, 74
Uyama, T., Xie, C., Aoyama, Y., et al. 2021, AJ, 162, 214
Venuti, L., Stelzer, B., Alcalá, J. M., et al. 2019, A&A, 632, A46
Wang, J. J., Ginzburg, S., Ren, B., et al. 2020, AJ, 159, 263
Wang, J. J., Vigan, A., Lacour, S., et al. 2021, AJ, 161, 148
Wilhelm, M. J. C., & Zwart, S. P. 2021, MNRAS, 509, 44
Zhou, Y., Bowler, B. P., Wagner, K. R., et al. 2021, AJ, 161, 244
Zhou, Y., Sanghi, A., Bowler, B. P., et al. 2022, ApJ, 934, L13

# Appendix A: UVES observations

The observing log is in Table A.1. Figures A.1– A.3 provide visual examples of the PSF-based spectral extraction process.

The interstellar extinction correction was obtained from the Leike & Enßlin (2019) 3D dust map, based on Gaia Data Release 2 (DR2). Using the MADYS toolkit (Squicciarini & Bonavita 2022), the extinction was integrated in a column out to the position of Delorme 1 (AB)b, resulting in $A_V = 0.009 \pm 0.003$ mag. This is a negligible contribution.

**Table A.1.** Observing log of the VLT/UVES 2021 October 25 observation of the Delorme 1 AB system during ESO programme 0108.C-0655(A).

| Date (MJD) | UT time (hh:mm) | Seeing ('') | Airmass |
|---|---|---|---|
| 59512.2338014 | 05:42 | 0.53 | 1.28 |
| 59512.2555725 | 06:08 | 0.52 | 1.34 |

**Notes.** (DIT, NDIT) of (1420 s, 1) for both integrations.

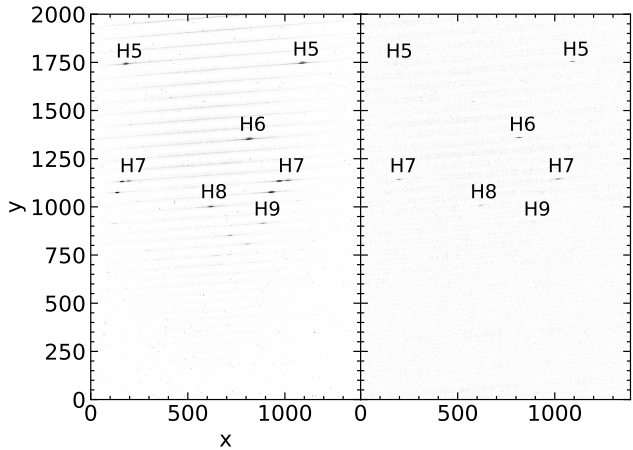

**Fig. A.1.** Full frame 1 reduced image before (left) and after (right) subtraction of the stellar lines. Confirmed H I detections are highlighted. The contrast in the image on the right was increased by a factor of five and hot pixels were masked for clarity.

## A.1. Pipeline reduction

Figures A.1 and A.3 show the full-frame image and other PSF extraction examples, respectively. Figure A.2 illustrates the reliability of the pipeline in accounting for cosmic rays.

The PSF subtraction was carried out through a process of iterative refining of the modelled stellar PSF. Following an initial basic reduction, the orders were first identified from the Flats. A column-by-column background per order was estimated based on the farthest few pixels from the order centre defined within a three-column radius, which was then subtracted from the data. Due to the strength of the companion lines in the raw data, the pixels at the companion location were masked out prior to PSF fitting for some orders, so that the core of the stellar PSF was accurately defined. Further, due to the bright continuum and the resulting high noise level in the data, the stellar PSF fit was performed in two iterations.

In the first iteration, a spline was fit to the $5\sigma$-clipped median of every 100 pixels in each order. The pixel data used for this pur-pose were first normalised using a windowed sum of ten pixels around the PSF core to avoid contamination due to noise or a companion signal. Once this first approximation of the PSF was obtained, a PSF frame was created based on a pixel-sampled PSF for each column. An initial 1D stellar spectrum was extracted from the frame, as well as a synthetic frame containing just the stellar signal computed from the PSF. A global re-normalisation was performed on this initial spectrum, and a second iteration of PSF fitting was then performed for each order by fitting a weighted spline to the whole order. Here, the data were normalised using the 1D spectrum from the previous step and the weights defined by the theoretical variance computed from the synthetic frame and the background. The resulting PSF was then used to create the final synthetic spectrum that was subtracted from the reduced data to get the first residual frame, containing only the companion signal (middle panels of Fig. A.3).

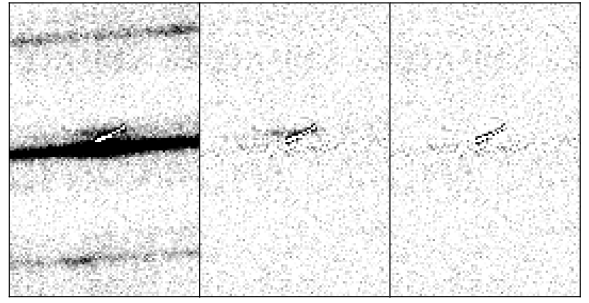

**Fig. A.2.** As in Fig. 1, but for H8 from frame 2, showcasing the reliability of the PSF-based subtraction for handling cosmic rays.

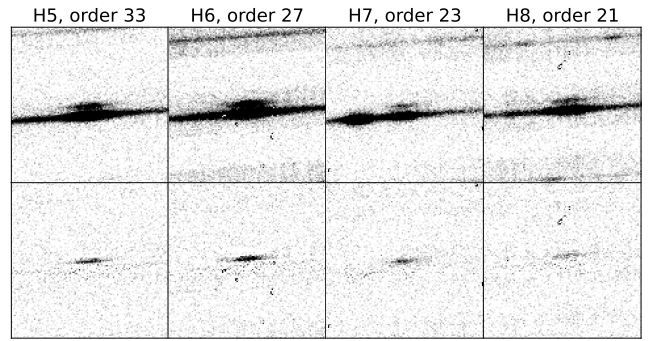

**Fig. A.3.** Mosaic showing cut-outs from frame 1 of H5–H8. Upper panels: Before subtraction of the stellar lines. Lower panels: After subtraction. Both panels show the lines at the same contrast, which is also used for the right panel in Fig. A.1.

The companion PSF frame was created using a spline fit (as above) on a slit with a radius of 8.7 pixels defined around the companion location. The 1D spectrum of the companion was extracted using this PSF frame and a resulting companion synthetic frame was then created. This was further subtracted from the first residual frame to give a second residual frame (bottom panels of Fig. A.3). This final residual frame has both the stellar and the companion signal efficiently extracted from the original data.

## A.2. Flux calibration

Flux calibration was done by calibrating the stellar binary flux using synthetic spectra obtained from the VO SED Analyser (VOSA; Bayo et al. 2008). The best fit to the available stellar photometry (excluding an IR excess from WISE) was the VOSA BT-Settl AGSS2009 model spectrum with $T_{eff}$ = 3000 K, $\log(g)$ = 5, and [Fe/H] = 0. The averaged stellar spectrum (of the two frames) was divided by the model spectrum after being interpolated to the same wavelength grid of a given spectral order. The correction for the blaze function curvature was then obtained from this divided spectrum with a spline fit. Finally, the correction was then applied on an order-by-order basis to each frame of the companion and stellar spectrum.

## Appendix B: Line modelling

### B.1. Gaussian fitting of H I emission lines

Here we detail the fitting procedure and tabulate the fitting parameters in Table B.1. Fig. B.1 compares the H7 line profiles from both orders. Plots of H5 are seen in Fig. B.2, H6 in Figs. B.3 and B.4, H7$_{23}$ in Fig. B.5, and H9 in Fig. B.6. Fitting was done iteratively using astropy (Robitaille et al. 2013; Astropy Collaboration 2018) and specutils (Earl et al. 2022) and evaluation of a fit was primarily done by eye, but also qualitatively through residuals and RMSE. It is worth noting that any Gaussian fitting, especially multiple fits, are degenerate, and only H7 in spectral order 24 could be readily fit with a double Gaussian without prior boundary constraints. To quantify the reliability of the fitted parameters better, uncertainties were estimated by simulating $10^3$ spectra of the observed averaged spectrum, through perturbation using a random normal distribution and the empirical uncertainties in flux. We note that H5 was detected in both spectral orders 33 and 34, and H7 in orders 23 and 24. However, H5$_{34}$ was badly affected by noise and only useful for extracting a similar RV shift in the line centre as H5$_{33}$.

In the stellar case, the width of the emission line, particularly $W_{10}$ of H$\alpha$, is related to $v_0$ and it can therefore be connected to an accretion rate (e.g. Natta et al. 2004). In the case of planetary accretion shock-heated emission, the width of the line is loosely coupled to the pre-shock $v_0$ and instead primarily determined by the effects of Doppler broadening (Aoyama et al. 2020). As such, we do not expect the line widths estimated in Table B.1 to correlate with the estimates of $v_0$ derived from the line profile analysis within the planetary accretion shock framework.

### B.2. Characterisation of tentatively detected emission lines

We describe the selection process for quantifying potential lines as tentative detections and list some parameters in Table B.2. Each order was normalised by its median flux, which excluded the order edges (first and last 100 pixels). Low confidence detections were assigned to lines above a normalised flux of five, if distinguished above the local noise floor. Conversely, the medium confidence level was set to a value of ten. If a low confidence line was also detected in both frames (or orders), it was given medium confidence. If it is also visible in the reduced data frame, the line is given high confidence. Potential emission lines (noted as 'Line?' in Table B.2) were identified using the NIST line database[1]. The works of Sicilia-Aguilar et al. (2012, 2015) were also used as examples of metallic lines detected during stellar accretion. Possible lines were primarily listed in Table B.2 for positive redshifts, and no potential blueshifted or redshifted forbidden transitions were identified to be nearby. It is possible that more lines are present near the average noise level, especially higher orders of H I. The tentative detection of Ca K (3933.66 Å) could not be characterised beyond the FWHM and RV shift as it is located in a very noisy region.

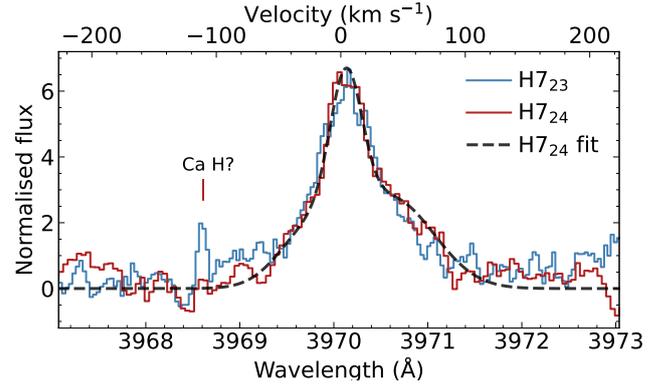

**Fig. B.1.** Figure showing the smoothed (moving-box average of three bins) normalised line profiles of H7 from orders 23 (blue) and order 24 (red). The Gaussian fit of order 24 of H7 (seen in Fig. 2) is indicated by the dashed black line. The tentative detection of Ca H in order 23 is highlighted.

We tentatively detect H 11 and H 12 with high and medium confidence. Furthermore, H11 is weakly visible in the data frames, but not as clearly as in Fig. B.7, while H12 cannot be reliably seen.

The tentative He I detections (4471.48 Å: $\log(L_{line}/L_\odot)$ = $-7.7 \pm 0.2$, 4026.19 Å: $\log(L_{line}/L_\odot)$ = $-8.2 \pm 0.2$ and 3888.64 Å) share a similar RV shift within a few km s$^{-1}$. They also differ slightly in FWHM, but are largely consistent within the uncertainties. Several He I lines were detected in the optical spectrum in Eriksson et al. (2020), but as these were completely unresolved, no constraints could be obtained for the line profiles. When feasible, we have estimated the accretion rates of tentative H and He I lines using the stellar scaling detailed in Sect. 3.1. We note that H 11 and H 12 estimates agree well with equivalent estimates for the confirmed detections (Table 1). In the stellar context, the estimates from He I differ by roughly an order of magnitude compared to Table 1, with an average of $\log(\dot{M}/M_J\,\mathrm{yr}^{-1}))$ = $-8.0 \pm 0.4$. Finally, the NIR spectrum from Betti et al. (2022) appears to show the detection of He I $\lambda$10830 Å. With helium lines having been detected from multiple epochs and instruments, this highlights the need for modelling of helium lines in a planetary accretion framework.

If the detection of Ca H is accurate with $\log(L_{line}/L_\odot)$ = $-8.8 \pm 0.2$, stellar scaling results in $\log(\dot{M}/M_J\,\mathrm{yr}^{-1}))$ = $-10.0 \pm 0.3$. This underestimates the accretion rate by more than an order of magnitude, compared to the H I lines. This could be related to the uncertainty in the calibration, the quality of the line, or indicate that the Ca H emission originates during different conditions in the planetary accretion paradigm.

---

[1] All in-air rest wavelengths reported in this Letter are from https://physics.nist.gov/PhysRefData/ASD/lines_form.html.

**Table B.1.** Delorme 1 (AB)b line profile characteristics of confirmed H I emission lines.

| Line | $\lambda_{\mathrm{rest,\,air}}$ (Å) | Fit [a] | $\lambda_{\mathrm{obs}}$ (Å) | RV (km s$^{-1}$) | FWHM (Å) | FWHM (km s$^{-1}$) | $W_{10}$ (km s$^{-1}$) |
|---|---|---|---|---|---|---|---|
| H5$_{33}$ (Hγ) | 4340.464 | d, NC | 4340.56 ± 0.04 | 7 ± 3 | 0.61 ± 0.06 | 42 ± 4 | 80 ± 8 |
| | | d, BC | 4340.59 ± 0.06 | 9 ± 4 | 1.41 ± 0.16 | 99 ± 11 | 184 ± 20 |
| H6 (Hδ) | 4101.734 | d, NC | 4101.84 ± 0.02 | 7 ± 3 | 0.54 ± 0.04 | 39 ± 3 | 77 ± 5 |
| | | d, BC | 4101.84 ± 0.02 | 7 ± 3 | 1.30 ± 0.08 | 95 ± 6 | 178 ± 9 |
| | | t, NC | 4101.83 ± 0.02 | 7 ± 2 | 0.52 ± 0.04 | 38 ± 3 | 66 ± 5 |
| | | t, BC$_1$ | 4102.13 ± 0.06 | 29 ± 6 | 1.09 ± 0.06 | 101 ± 6 | 149 ± 11 |
| | | t, BC$_2$ | 4101.40 ± 0.07 | −25 ± 6 | 0.93 ± 0.09 | 68 ± 7 | 178 ± 8 |
| H7$_{24}$ (Hε) | 3970.072 | d, NC | 3970.10 ± 0.06 | 3 ± 4 | 0.38 ± 0.07 | 29 ± 6 | 49 ± 10 |
| | | d, BC | 3970.34 ± 0.08 | 20 ± 6 | 1.36 ± 0.10 | 103 ± 7 | 184 ± 17 |
| | | t, NC | 3970.13 ± 0.08 | 5 ± 5 | 0.40 ± 0.05 | 30 ± 5 | 43 ± 7 |
| | | t, BC$_1$ | 3970.62 ± 0.11 | 40 ± 8 | 1.06 ± 0.13 | 80 ± 10 | 147 ± 17 |
| | | t, BC$_2$ | 3969.77 ± 0.10 | −22 ± 7 | 0.82 ± 0.06 | 62 ± 7 | 115 ± 12 |
| H7$_{23}$ (Hε) | 3970.072 | t, NC | 3970.12 ± 0.05 | 4 ± 4 | 0.54 ± 0.07 | 41 ± 4 | 68 ± 9 |
| | | t, BC | 3970.32 ± 0.06 | 19 ± 4 | 1.60 ± 0.12 | 124 ± 8 | 226 ± 19 |
| H8 | 3889.049 | t, NC | 3889.09 ± 0.07 | 5 ± 5 | 0.60 ± 0.06 | 45 ± 5 | 84 ± 9 |
| | | t, BC | 3889.30 ± 0.09 | 23 ± 7 | 1.53 ± 0.18 | 118 ± 12 | 218 ± 25 |
| H9 | 3835.384 | s | 3835.57 ± 0.03 | 14 ± 2 | 0.58 ± 0.08 | 45 ± 7 | 82 ± 12 |

**Notes.** [a] Single (s), double (d), or triple (t) Gaussian fits of narrow (NC) and broad components (BC). The triple fits for H7$_{23}$ and H8 include the tentative Ca H and He I lines, respectively (see Table B.2).

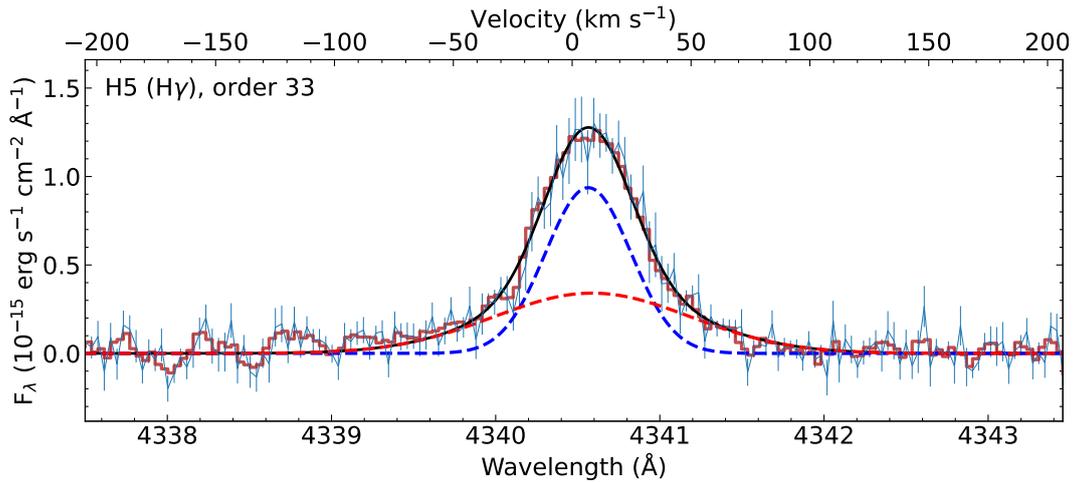

**Fig. B.2.** Same as in Fig. 2, but showing the best double Gaussian fit for H5.

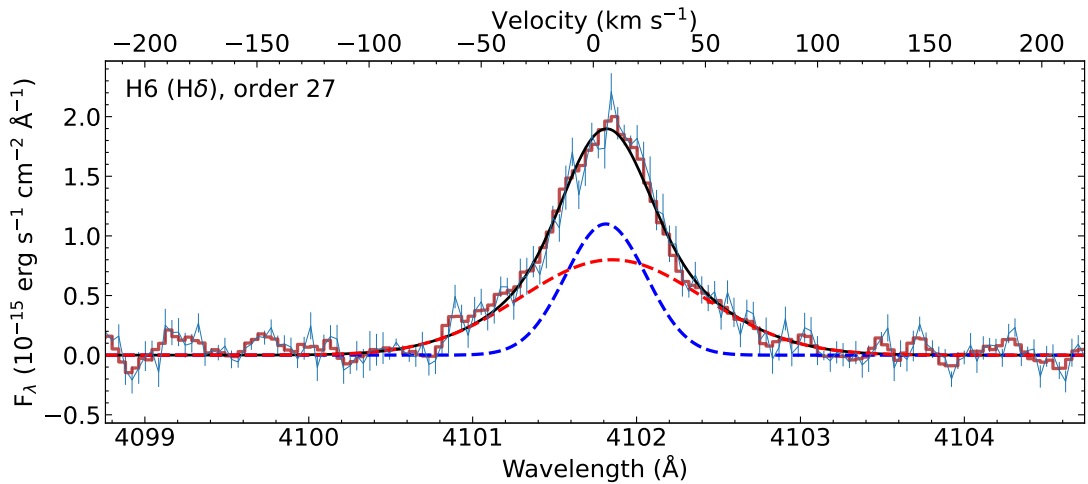

**Fig. B.3.** Best double Gaussian fit for H6.

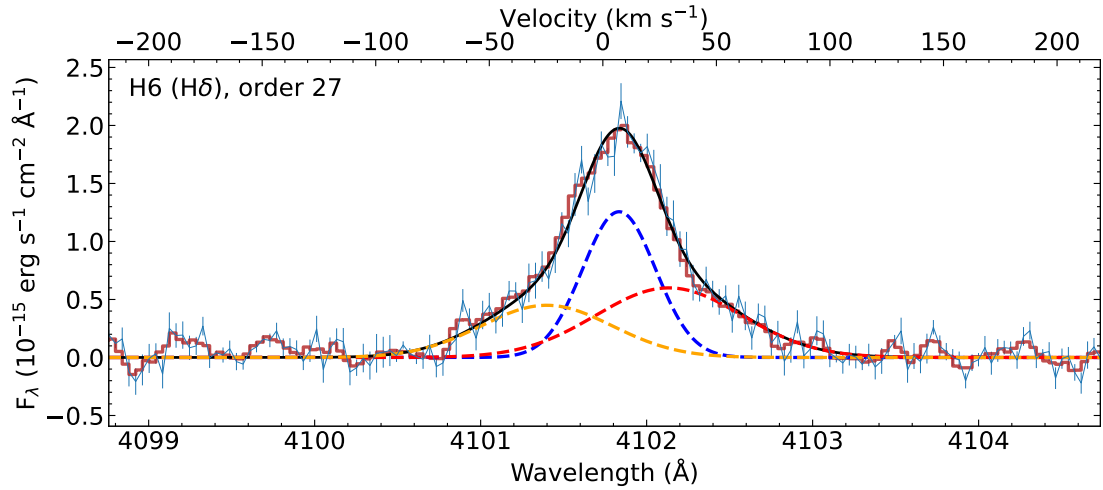

**Fig. B.4.** Three-component fit of H6 using one NC and two BCs, illustrating that the line asymmetries in H5–H8 can also be reasonably fit when a blueshifted BC is included, as discussed in Sect. 4.

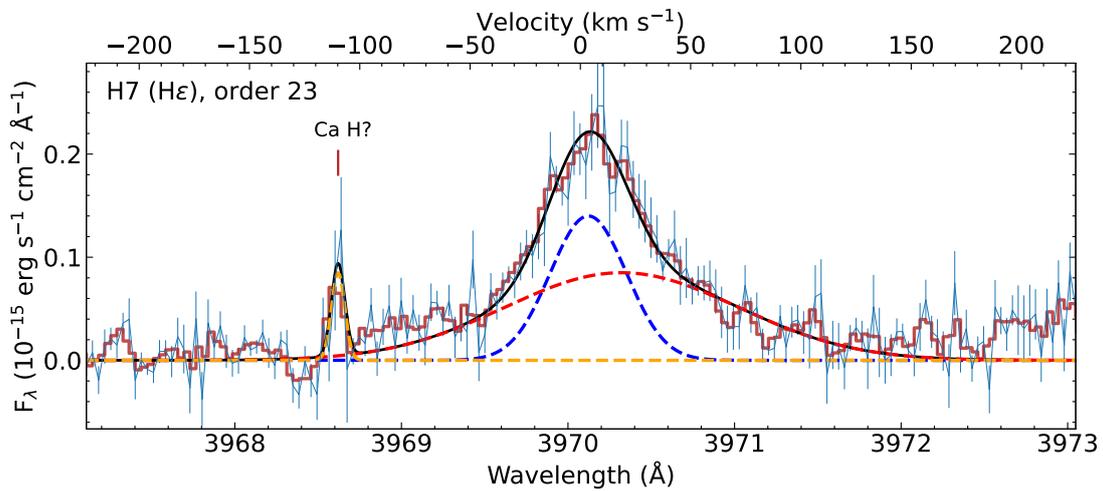

**Fig. B.5.** Best fit for H7$_{23}$, including a possible contribution from the Ca H line.

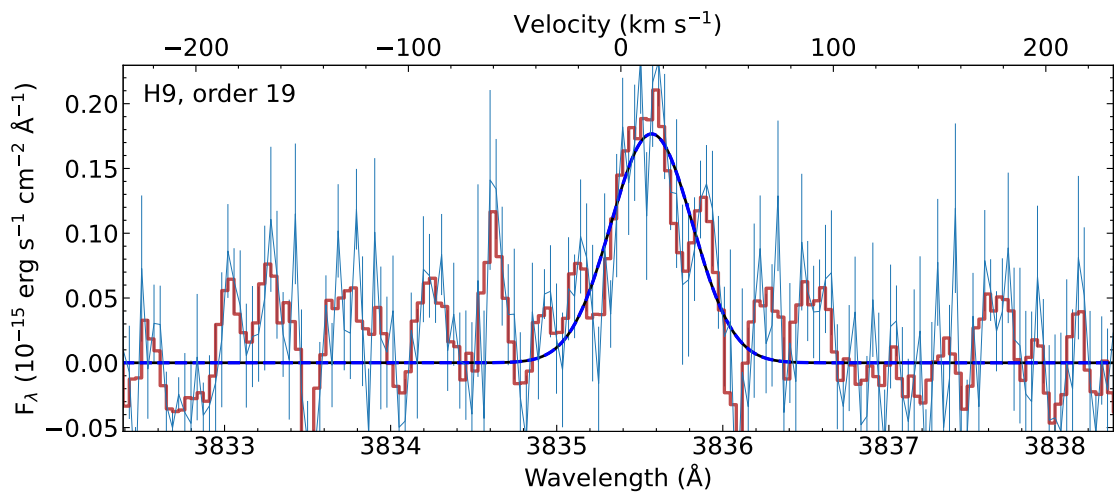

**Fig. B.6.** Best fit for H9. The low S/N prevented the reliable fitting of multiple components.

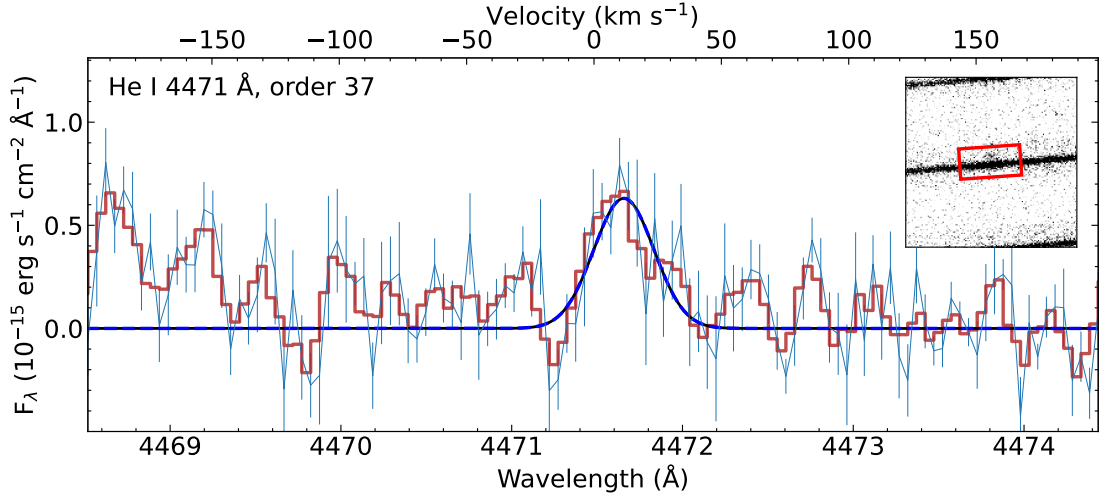

**Fig. B.7.** High-confidence tentative detection of He I λ4471.66. The inset covers the same wavelength window of frame 1 and shows the detection above the broad stellar continuum. The red rectangle corresponds to a 2-Å window around the line centre.

**Table B.2.** Delorme 1 (AB)b line characteristics for tentatively detected emission lines.

| $\lambda_{obs}$ [a] (Å) | FWHM (Å) | FWHM (km s$^{-1}$) | Line? | $\lambda_{rest, air}$ (Å) | RV (km s$^{-1}$) |
|---|---|---|---|---|---|
| *High confidence* | | | | | |
| $4471.63_{37}$ | $0.40 \pm 0.05$ | $28 \pm 3$ | He I | 4471.4802 | $12 \pm 3$ |
| $4026.34_{25}$ | $0.25 \pm 0.05$ | $20 \pm 4$ | He I | 4026.1914 | $9 \pm 3$ |
| $3770.18_{17}$ | $0.41 \pm 0.11$ | $32 \pm 8$ | H11 | 3770.063 | $9 \pm 5$ |
| *Medium confidence* | | | | | |
| $4023.80_{25}$ | 0.11 | 9 | Cr I | 4023.7336 | 5.0 |
| | | | Ti I | 4023.6789 | 8.9 |
| | | | Cr I | 4023.43 | 27.6 |
| $4017.17_{25}$ | 0.12 | 8 | Fe I | 4017.1483 | 1.6 |
| | | | Fe I | 4017.0832 | 6.5 |
| | | | Ti I | 4016.9636 | 15.6 |
| | | | Cr I | 4016.8961 | 20.4 |
| $3968.62_{23}$ | $0.09 \pm 0.02$ | $7 \pm 2$ | Ca II H | 3968.47 | $11 \pm 2$ |
| $3949.90_{23}$ | 0.14 | 11 | Cr I | 3949.6132 | 18.6 |
| $3933.76_{22}$ | 0.12 | 9 | Ca II K | 3933.66 | 7.6 |
| $3888.77_{21}$ [b] | $0.26 \pm 0.08$ | $23 \pm 6$ | He I | 3888.648 | $7 \pm 4$ |
| $3887.19_{21}$ | 0.06 | 5 | Cr I | 3887.08 | 8.5 |
| | | | Fe I | 3887.048 | 11.0 |
| | | | Cr I | 3886.7953 | 30.4 |
| $3866.51_{20}$ | 0.09 | 7 | Ti I | 3866.4394 | 5.5 |
| $3750.40_{16}$ | $0.26 \pm 0.07$ | $21 \pm 5$ | H12 | 3750.151 | $20 \pm 3$ |
| *Low confidence* | | | | | |
| $4465.97_{37}$ | 0.13 | 9 | Ti I | 4465.806 | 11.0 |
| $4421.56_{37}$ | 0.12 | 8 | Ti I | 4421.451 | 7.4 |
| $4258.58_{31}$ | 0.19 | 13 | Ti I | 4258.525 | 3.9 |
| | | | Fe I | 4258.3156 | 18.6 |
| $4245.43_{31}$ | 0.22 | 17 | Fe I | 4245.3441 | 6.1 |
| | | | Fe I | 4245.2569 | 12.2 |
| $3896.23_{22}$ | 0.16 | 12 | Cr I | 3896.01 | 17.3 |

**Notes.** [a] Spectral order noted in subscript. [b] From fit of blended H8 (Fig. 2). Not included in the relations by Alcalá et al. (2017).

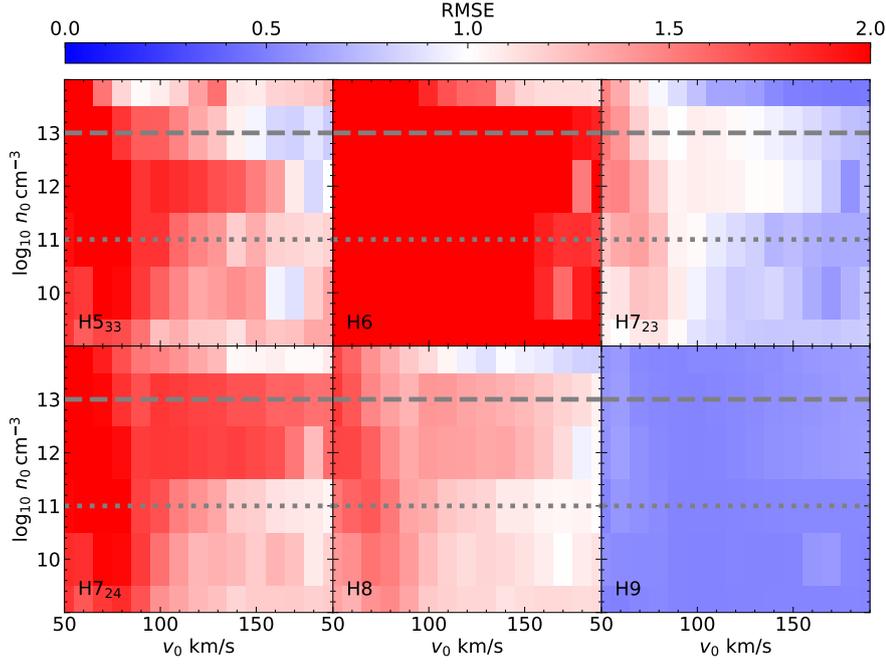

**Fig. B.8.** Root mean square error (RMSE; common colour scale on top), for the analysis described in Sect. 3.2, using the shock models of Aoyama et al. (2018, 2020, 2021). The two input parameters are the pre-shock gas velocity $v_0$ and the number density of hydrogen nuclei $n_0$. The smaller the RMSE is, the better the fit with the one-sigma regions (RMSE $\lesssim 1$; Eq. (3)) in blue. Upper panels: H5$_{33}$, H6, and H7$_{23}$. Lower panels: H7$_{24}$, H8, and H9. The dashed and dotted horizontal lines indicate $n_0 = 10^{13}$ cm$^{-3}$ and $n_0 = 10^{11}$ cm$^{-3}$, respectively. At higher densities ($n_0 \gtrsim 10^{13}$ cm$^{-3}$), both the NC and BC of the data are reproduced by the model, while the lower density only accurately reproduces the NC. The lack of constraints on the parameters from the H9 line is likely due to the low line-S/N.